# Neural network setups for a precise detection of the many-body localization transition: Finite-size scaling and limitations

Hugo Théveniaut and Fabien Alet

*Laboratoire de Physique Théorique, IRSAMC, Université de Toulouse, CNRS, UPS, 31062 Toulouse, France*

Determining phase diagrams and phase transitions semiautomatically using machine learning has received a lot of attention recently, with results in good agreement with more conventional approaches in most cases. When it comes to more quantitative predictions, such as the identification of universality class or precise determination of critical points, the task is more challenging. As an exacting testbed, we study the Heisenberg spin-1/2 chain in a random external field that is known to display a transition from a many-body localized to a thermalizing regime, which nature is not entirely characterized. We introduce different neural network structures and dataset setups to achieve a finite-size scaling analysis with the least possible physical bias (no assumed knowledge on the phase transition and directly inputting wave-function coefficients), using state-of-the-art input data simulating chains of sizes up to $L = 24$. In particular, we use domain adversarial techniques to ensure that the network learns scale-invariant features. We find a variability of the output results with respect to network and training parameters, resulting in relatively large uncertainties on final estimates of critical point and correlation length exponent which tend to be larger than the values obtained from conventional approaches. We put the emphasis on interpretability throughout the paper and discuss what the network appears to learn for the various used architectures. Our findings show that a *quantitative* analysis of phase transitions of unknown nature remains a difficult task with neural networks when using the minimally engineered physical input.

## I. INTRODUCTION

The recent application of machine learning techniques to condensed matter and statistical physics led to several and important successes in various problems, ranging from the detection of phases of matter from synthetic [1–9] or experimental data [10,11], wave-function reconstruction [12], the improvement of variational *Ansätze* for quantum problems [13–18], and efficient Monte Carlo sampling [19–21], in such a way that machine learning is now regarded as a new tool for the study of complex, interacting, (quantum) physical systems [22–24].

In the case of phase identification, the semiautomatic discovery of phase transitions and mapping of phase diagrams rely on the ability of machine learning algorithms to extract the relevant features for the classification of samples from large datasets. Data consist for instance of Monte Carlo snapshots of configurations or measurements of various types of observables. This approach has enabled the recovery of known phase diagrams or the location of phase transitions with qualitative agreement with more conventional approaches (based for instance on order parameters, and/or theory of finite-size scaling), achieving this *at a much lower computational cost*, e.g., using fewer samples or smaller system sizes. In some cases, critical exponents have been extracted by an analysis of the neural networks (NN) outputs [1,25–28], a particularly nontrivial prediction.

There are cases however where machine learning techniques fail to capture the correct physical behavior, or at least not as correctly as the conventional approaches do [3,6,7,29,30]. The selection of the input data is paramount in this method, as providing engineered quantities known to have a physical content naturally helps NN to be more accurate using even more modest resources (input data or network size) [3,31]. It can indeed sometimes require a bit of manual feature engineering to accurately grasp the physical behavior in a transition region [3,29,31].

A natural question is whether machine learning can lead to superior results in the case of unknown phase diagrams or for systems where conventional approaches have difficulties. One speaking modern example, and the focus of the current work, is the many-body localization (MBL) transition in one dimensional quantum disordered systems. There, finite-size effects are crucial to apprehend the transition as the size of available samples is limited (contrary to classical or quantum problems that can be treated with Monte Carlo simulations). There is furthermore no accepted finite-size scaling theory for this transition. Conventional approaches based on the extensive study of various physical quantities provide an estimate of the phase transition [32], but may be hampered by finite-size effects (for the standard MBL model considered in this work, the maximum sample sizes that can be used to probe the transition regime is $L = 24$ [33]). For instance, attempts at performing finite-size scaling [32,34] result in a critical exponent for the correlation/localization length which does not match predictions from renormalization group approaches and do not fulfill a bound argued to be valid for the MBL phase transition.

In this work, we provide a detailed analysis of a neural network, from its construction to the treatment of its output,

designed to locate the MBL phase transition in a prototypical 1D quantum model. Our goal is *not* to engineer the best network architecture that reproduces the known estimate of the phase transition with the smallest amount of input data, but rather to see if we can go beyond by using the same (high) quality of input data. In short, we ask the following question: can a NN approach provide a quantitative description of the transition (not just qualitative), and in particular, improve determination of critical point and exponents? In doing so and as a probe of the efficiency of this approach for unknown phase diagrams, we furthermore wish to provide the least engineered input.

There have been several prior works that used NN approaches to locate the MBL phase transition in 1D disordered quantum systems. One group of works [2,35–39] considered inputing the *entanglement spectrum* of eigenstates, resulting in phase diagrams in (qualitative) agreement with conventional approaches. One notes however that the entanglement spectrum is a high-level engineered quantity where physical features are already extracted (as for instance the two phases around the MBL transition have a different scaling behavior for the entanglement entropy), and that there was no systematic study of finite-size effects on the prediction of the network. Other works [40,41] considered locating the phase transition using information from dynamical measurements (such as time traces of observables after a quench). While larger systems can be used using this approach, finite-time effects (contrary to eigenstates which are probes of infinite-time behavior) may be relevant, especially close to the transition. The question of which observable to input also leaves more room for feature engineering in this case. Nevertheless, this approach may be particularly relevant for experiments which probe the MBL transition [42–45], and which are precisely based on measurements of finite-time traces after a quench. Finally, two recent works [26,27] considered inputing directly the eigenfunctions in order to detect the MBL transition, this time supplemented by a finite-size scaling analysis.

In this work, we follow this last approach of inputting the wave function, as this is likely the best approach to provide unbiased information to the network (see discussion below in Sec. III A). We provide a detailed analysis of the influence of the network architecture and hyperparameters to the predictions. Quite crucially, we consider input data obtained from large system sizes up to $L = 24$ spins, that is at, or beyond, the state-of-the-art numerics used with conventional approaches. The plan of the manuscript is as follows. Section II describes the lattice model used in this study and briefly recapitulates aspects of its MBL transition. In Sec. III, we provide an extensive description on possible NN setups to study the transition, discussing the choice of input data (Sec. III A), network architectures (Sec. III B), and hyperparameters (Sec. III C), as well as output treatment (Sec. III D). The remainder of the paper presents our results using three different setups: a single-size training setup where an ensemble of NN are separately trained on different system sizes (Sec. IV), a multisize training setup where one NN is trained on a dataset containing data from multiple system sizes all at once (Sec. V) including a constraint in the form of a domain adversarial component to achieve better generalization (Sec. VI). Section VII critically discusses these results and summarizes the open questions and challenges for the detection of MBL physics using NN.

## II. MODEL AND MBL TRANSITION

Many-body localization (see Refs. [46–50] for introduction and reviews) is an active research area which aims at understanding the possibility of survival of Anderson localization in many-body strongly interacting quantum systems. Existence of the transition to a MBL phase is now accepted for one-dimensional lattice models in the presence of strong-enough disorder. The hallmarks of MBL include low-entanglement in eigenstates (even in the middle of the many-body spectrum), absence of thermalization (the eigenstate thermalization hypothesis [51,52] is not respected) and validity of thermodynamic ensemble, emergence of integrability (through the form of local integrals of motions), memory of initial conditions in quench setups, etc. [46–50]. All these specificities have been used as probes of the existence of a MBL phase in various studies, however here we would like not to impose the use of any specific probe but use machine learning to detect the transition to MBL.

We perform computations on the standard lattice model of MBL, namely, the spin 1/2 Heisenberg chain with disorder [53,54]:

$$H = \sum_{i=1}^{L} \mathbf{S}_i \cdot \mathbf{S}_{i+1} - \sum_{i=1}^{L} h_i S_i^z. \quad (1)$$

Here, $\mathbf{S} = (S^x, S^y, S^z)$ denotes a vector of spin-1/2 operators. This model has been shown to display a phase transition between a low-disorder ergodic phase which satisfies the eigenstate thermalization hypothesis (ETH), and a high-disorder MBL phase [54]. The transition line depends on the energy density of the eigenstates, forming a many-body mobility edge [32]. In this work, we will mainly consider eigenstates at "infinite temperature" (in the middle of the spectrum), corresponding to a normalized energy density $\epsilon = (E - E_{\min})/(E_{\max} - E_{\min}) = 0.5$, where $E_{\min/\max}$ are the energy spectrum extrema. For $\epsilon = 0.5$, the transition has been estimated to occur at $h_c \simeq 3.7$ in Ref. [32] through the use of various physical estimators. The nature of the MBL phase transition and the possibly associated critical exponents, is still a challenging open question. Assuming a standard second-order phase transition and scaling behavior with a physical length scale (correlation or localization length) diverging close to the transition as $\xi \sim |h - h_c|^{-\nu}$, the finite-size analysis of Ref. [32] estimated a value $\nu \simeq 0.8, 1.0$ for most physical observables considered (using system sizes up to $L = 22$). On the other hand, renormalization group analysis based on effective simplified classical models predict a much larger value $\nu \simeq 3.5$ [55–57] or even $\nu = \infty$ (Kosterlitz-Thouless scenario advocated in recent works [58–60]). Also, the Harris-Chayes criterion which has been argued [61] to hold also for the MBL transition provides a bound $\nu \geqslant 2$ in dimension 1. It was also pointed out that the finite-size effects could be particularly important for this model with random disorder [62].

## III. BUILDING A NEURAL NETWORK TO STUDY THE MBL TRANSITION

There are many possible ways to design a neural network aimed at detecting the MBL transition, which can vary from the choice of input data, architecture and hyperparameters of the network, as well as the interpretation of the network output. In order to perform these choices, we are guided by the following principles: (i) minimal manual feature engineering: that is, we want the input data not to be preprocessed with already-extracted physical features that could bias the predictions, (ii) scalability: the network should be able to treat data from different system sizes (in order to perform finite-size scaling), (iii) keep variability with respect to irrelevant (unphysical) parameters as small as possible, and (iv) interpretability: the architecture should allow for possible physical explanations of what the machine actually learned. In the following, the interpretation of the neural network will be achieved by an analysis of its internal weights.

### A. Choice of input data

The choice of input data and its formatting is of major importance in the context of detecting phase transition with supervised learning. Indeed, this method relies on the fact that the NN will be able to learn the relevant phase characteristics being trained only in two extreme limits of the phase diagram.

In this work, we directly input the eigenstates. One caveat is that this requires to specify a basis set in which to expand. Given an eigenstate $|n\rangle$, we expand it in the $S^z$ computational basis $|i\rangle$: $|n\rangle = \sum_i c_i |i\rangle$ and we denote $p_i \equiv |c_i|^2$. The choice of this basis stems from the fact that the $S^z$ basis diagonalizes the model (1) in the infinite disorder limit. Moreover, basis-dependent quantities such as the inverse participation ratios $\text{IPR}(|n\rangle) = \sum_i p_i^2$ or associated participation entropies $S_q^P(|n\rangle) = \frac{1}{1-q} \log \sum_i p_i^q$ have been shown to capture different behaviors in the two phases [32,33,63]. At the technical level, this is also the basis where the largest system sizes can be studied (as the Hamiltonian is quite sparse).

However, the exponentially growing number of coefficients with system size will eventually lead to computational issues for the largest systems, e.g., for $L = 24$ each eigenstate has more than 2.8 millions coefficients, which when multiplied by the number of eigenstates per disorder realization and the number of disorder realizations amounts to an extremely large amount of input data. This would entail very slow training and necessary lossy compression implemented in the NN architecture with many pooling layers for instance. We chose to engineer this compression step by hand: our solution is to keep only the largest coefficients of each eigenstate, we truncate the $N_c$ largest $p_i$. For illustration, we present in Fig. 1 typical $p_i$s for two values of disorder representative of the ETH and MBL phases. Note that the basis states associated to the largest coefficients differ from one eigenstate to another. The fact that the input data are now of fixed size (independent of $L$) will be useful when training with data from multiple sizes at once in Sec. V.

A lot of information is certainly lost in doing so, but we argue that it may not be crucial. On the MBL side, the local integral of motions picture [64–67] indicates that eigenstates have all typically the same structure coming from the strong disorder limit, with a very strong coefficient for a particular $S^z$ basis state (different for each eigenstate). On the ETH side, from random matrix theory, one expects a random coefficient structure with no correlation between basis states. As a further argument, one notes that IPR and participation entropies $S_q^P$ (for $q \geqslant 1$) are dominated by large $p_i$ (independently of which basis state corresponds to index $i$, one eigenstate from another).

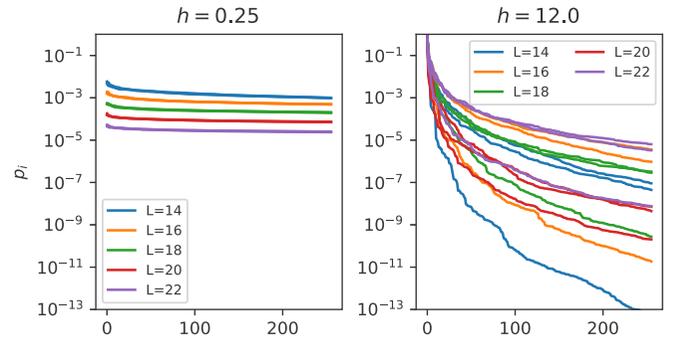

FIG. 1. Examples of $N_c = 256$ highest probabilities $p_i$ for eigenstates in the middle of the spectrum ($\epsilon = 0.5$) for different disorder realizations and system sizes for two disorder values located strongly in the ETH (left) and MBL (right) phases.

Data normalization is often recommended in conventional machine learning applications [68], as it helps a lot accelerating or even rendering possible the learning process. In our context, we want to avoid as much as possible this feature engineering step since it can bias the data and possibly lead to false estimate of the transition point. As an example, if we kept the sample size constant from one system size to another not by truncating but down-sampling [69] the largest probabilities, this would eventually lead to an underestimation of the critical disorder because down-sampling would lead to a faster decay of the highest probabilities eventually making the eigenstates look more "MBL" than they are.

Finally, we note that methods based on sampling of the eigenstates (such as quantum Monte Carlo) will pick up the basis states precisely with a probability $p_i$, and therefore this choice of truncation may be useful in other contexts where the exact eigenstates cannot be reached.

We obtain exact eigenstates at $\epsilon = 0.5$ of model (1) with the shift-invert method [70]. We use periodic boundary conditions, and consider eigenstates in the $S^z = 0$ sector (the total magnetization $S^z = \sum_r S_r^z$ is conserved in this model). We insist on having a large, state-of-the-art dataset. For training, we use 1000 realizations of disorder per disorder strength and 250 (respectively, about 150) realizations of disorder at prediction time for sizes $L = 14$, 16, 18, 20, 22 (respectively, $L = 24$). For each realization of disorder, we compute 100 (respectively, 60) eigenstates for $L \leqslant 22$ (respectively, for $L = 24$). We use a fine grid of disorder strength, specially close to the alleged transition region.

We present results obtained when providing the $N_c$ largest probabilities $p_i$ (Secs. IV–VI). In Appendix C, we furthermore consider the coefficients $c_i$ (i.e., restoring the sign) of the largest $N_c$ amplitudes as inputs for the neural networks.

### B. Choice of neural network architecture

As can be seen in Fig. 1, strongly ETH and MBL samples are in fact linearly separable (a threshold value for the largest $p_i$ suffices), hence the use of a neural network for this classification task appears unnecessary at first sight. However, our actual task is to not only perform well on the well-defined labeled region of the phase space but more importantly to assign labels to samples in the transition region. Therefore we can view our work as a benchmark of the NN ability to capture the relevant features and finite-size trends of the MBL transition, or put another way whether NN are a good *Ansatz* for the classification of phases present in model (1).

The chosen neural-network architecture is simple to keep its interpretation possible to a reasonable extent, and its optimization is standard. Nevertheless, we provide details for clarity purposes (we also refer the interested readers to Ref. [71] for an introduction to machine learning for physicists).

Artificial neural networks and in particular fully connected feed-forward neural networks are based on elementary units called artificial neurons. These units are simple functions that take a vector **x** of real values and transform it according to

$$y = g(\mathbf{W} \cdot \mathbf{x} + b), \quad (2)$$

where $g$ is a nonlinear so-called *activation* function, **W** is a vector of weights and $b$ a real weight called *bias*. Similarly, one can define a layer of artificial neurons which implements a mapping between an input vector **x** and an output vector **y** as follows:

$$\mathbf{y} = g(\hat{\mathbf{W}} \cdot \mathbf{x} + \mathbf{b}), \quad (3)$$

where $\hat{\mathbf{W}}$ is now a matrix of weights and **b** a vector and $g$ is applied element-wise on the input vector. This way, one can successively stack layers of artificial neurons, building more and more complex functions. There is some flexibility in the choice of $g$ and frequently used activation functions include

$$\text{ReLU}(x) = \begin{cases} 0 & \text{if } x < 0 \\ x & \text{if } x \geq 0 \end{cases}, \quad (4)$$

$$\text{ELU}(x) = \begin{cases} e^x - 1 & \text{if } x < 0 \\ x & \text{if } x \geq 0 \end{cases}, \quad (5)$$

$$\text{Softmax}_i(x) = \frac{e^{-x_i}}{\sum_j e^{-x_j}}. \quad (6)$$

Since our task is to classify eigenstates as being ETH or MBL, our neural network is a function that takes as an input an eigenstate in the form of a vector of size $N_c$ and outputs its label, that is (0,1) if it is ETH and (1,0) if MBL. The network used in the following is shown schematically in Fig. 2, there is one hidden layer of 32 neurons with $g_1 =$ ELU activation functions [see Eq. (5)] and a two-neuron output layer with $g_2 =$ Softmax activation function [see Eq. (6)]. Due to the softmax activation function, the output vector of the neural network is real and normalized to 1, thus it can be interpreted as probabilities to belong either to the ETH or the MBL phase. Training is done through stochastic gradient descent of a cross-entropy cost function with ADAM optimizer [72].

For our NN architecture and for the input data chosen, we found that the usual choice of ReLU activation functions (4) actually induces a bias in the location of the phase transition through the appearance of dead neurons that limit the NN capacity. This effect is described in length in Appendix A. The problem is avoided using other activation functions like tanh, leaky rectified linear unit (ReLU), or exponential linear units (ELU) [73], the latter being used in this work.

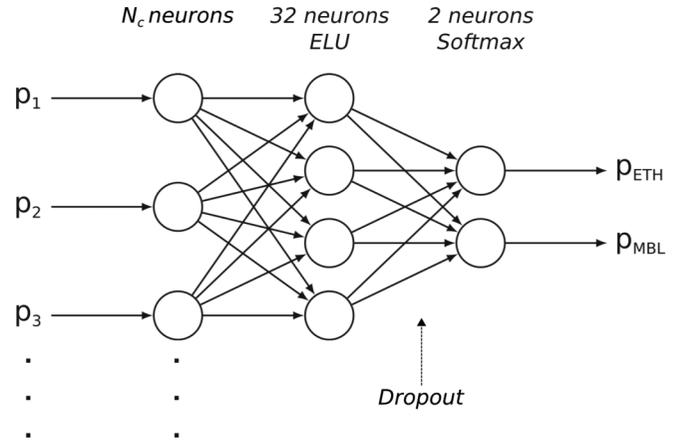

FIG. 2. Neural-network architecture used in this work.

When used along with ELU units, we noticed that dropout [74] brings additional practical benefits. This regularization technique consists in randomly dropping connections between neurons (here between the hidden layer and the output layer) during training. This prevents neurons from coadaptating and allows to learn feature detectors that are indeed more independent of each other.

### C. Model selection

In most traditional classification problems, model selection is done with respect to predictions on a labeled test set. For instance, a low test accuracy reveals that the model is unable to generalize well to unseen samples. In our case, all considered neural networks achieved 100% accuracy on training and test sets, but this only says that our data and chosen architecture are extremely good at distinguishing strongly ETH from strongly MBL samples. Given that our actual task is to assign labels to samples from the transition region, we need to find other ways of discriminating the NN performance.

One possibility is to ensure that the learned model achieves low bias and low variance. On the one hand, we argue that bias is low having checked that increasing the number of hidden neurons does not change the predictions, rather increasing variance. On the other hand, variance is kept small by choosing a relatively small number (32) of hidden neurons. Moreover, we can track the variance using cross-validation, i.e., obtaining multiple training instances from random initialization of the NN weights and random partitioning of the training datasets (as we leave aside a fraction of the data in a separate test set). In most cases, we observe a low and stable (during training) variance with a learning rate empirically chosen at $\alpha = 0.01$, batch size of $N = 1000$ samples and number of epochs of 5. The variance gets problematic when an adversarial component is added and further comments are provided in the corresponding Sec. VI.

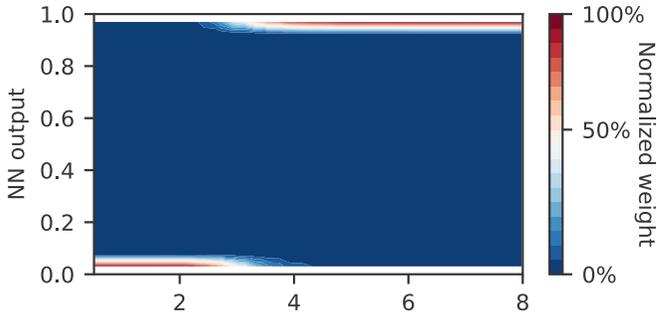

FIG. 3. Color histogram of the output of an exemplary neural network trained with $L = 18$ data evaluated on 300 disorder realizations for each disorder strength.

### D. Output analysis

As can be seen in Fig. 3, we observed that the typical distribution of the NN output for a given system size $L$ and disorder strength $h$ is unimodal, the distribution having very low variance around 0 (1) in the ETH (MBL) phase. This motivates the choice of considering the fraction $f$ of samples whose classification confidence is above 0.5 as a good quantity to faithfully describe the output of the neural network. Note that $f$ is then the proportion of MBL-classified eigenstates from an ensemble of eigenstates coming from different disorder realizations and classified by different training instances. We will clarify later how we compute the error bars on this quantity (see Appendix B for more details on the different sources of classification variance).

To the best of our knowledge, there is no theory which describes the finite-size scaling (with $L$) of the network output. Indeed, there is in general no expectation for which kind of physical observable (if any) the output will correspond to: for a standard continuous phase transition, $f$ could for instance mimic the order parameter or its Binder cumulant (or any combination thereof), which are known to display different critical behavior and finite-size effects. Various phenomenological scalings have thus been used in the literature. When output curves for different $L$ cross as a function of the control parameter $h$, a natural scaling form is $f = g[L^{1/\nu}(h - h_c)]$, with $h_c$ the critical disorder strength and $\nu$ the exponent associated to the divergence of a correlation/localization length $\xi \sim |h - h_c|^{-\nu}$. This is the form that was used, e.g., in Ref. [1] for the Ising model, or for the MBL transition in Ref. [27]. When curves for $f$ do not cross, one can alternatively try to define a finite-size pseudo critical point $h_c(L)$ (with some criterion) and naturally assume a finite-size relation $h_c(L) - h_c \sim L^{-1/\nu}$. This was for instance used in Refs. [25,26]. In our case, we find (see Figs. 4 and 6) that the latter situation applies (no crossing of curves) and thus assume the second scaling form.

In this case, there is a variety of options for the definition of $h_c(L)$ as can be seen in earlier works: one can introduce a confidence threshold $p_c$ as performed in Ref. [35] and look for maximum of the confusion curves, alternatively one can pinpoint the transition when the mean output curves reach 0.5 as in Ref. [26], or also consider the maximum of the confusion as defined in Ref. [40]. In the following, we define $h_c(L)$ to be the disorder strength at which half of samples are classified as MBL, meaning $f(h_c(L)) = 0.5$.

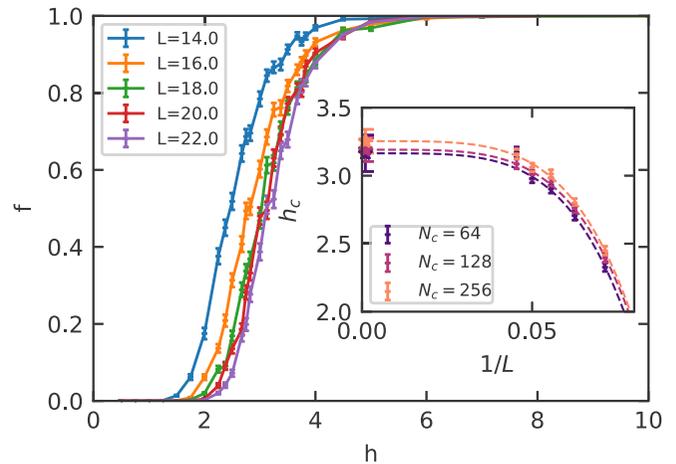

FIG. 4. Fraction of MBL-classified samples as a function of disorder strength for NN trained on a given system size $L$. Predictions are averaged over 250 disorder realizations per disorder (with 100 eigenstates per realization) and 50 training instances. Truncation order is $N_c = 256$. The error bars indicate the statistical error due to sampling disorder realizations. (Inset) Finite-size scaling analysis from $h_c(L)$ defined as $f(h_c(L)) = 0.5$ for different truncations $N_c$. the error bars on the final estimates come from the fitting procedure.

## IV. SINGLE SYSTEM SIZE TRAINING

The most direct way to do a finite-size study of model (1) assisted by neural networks is to train one NN for each system size. Hence, we study the predictions of five neural networks trained on data respectively from $L = 14, 16, 18, 20, 22$. Apart from the training dataset, all hyperparameters (learning rate, batch size, number of epochs, etc.) and NN architecture (number of hidden neurons, etc.) are fixed. The training dataset consists of eigenstates obtained at $h = 0.25$ (respectively, $h = 12.0$) for ETH (respectively, MBL)-labelled samples for all system sizes.

Figure 4 shows the fraction $f$ of MBL-classified samples, i.e., if $y_{\theta,r,i}(h)$ denotes the probability of eigenstate $i$ from disorder realization $r$ of being classified as MBL by the neural network $\theta$, then $f(h) = \frac{1}{N_\theta N_r N_i} \sum_{\theta,r,i} \Theta(y_{\theta,r,i}(h) - \frac{1}{2})$, where $\Theta$ is the Heaviside step function. As eigenvectors of the same disorder realization are correlated and the neural networks have very low variance, we chose to bin quantities over all eigenstates of the same realization and all neural networks, and then compute the standard error over these bin averages (as performed in Ref. [32]), in order not to underestimate error bars. Appendix B gives further details on the variations of sample classification from one NN instance to another, including a discussion on predictions for individual eigenstates and their correlation with entanglement entropy.

Several features can be distinguished: one is the existence of a fully ETH region (where all samples are classified as ETH) that extends from $h = 0$ to 2 and a fully MBL region starting from $h = 6$ for all system sizes. Another distinct feature is the hierarchy of the curves depending on the system size $L$, i.e., the crossover from ETH to MBL happens for higher disorder as $L$ is increased. This behavior is in agreement with many other observables (such as spectral statistics, entanglement variance, dynamical spin fraction) used in the standard analysis of this system [32], which also display regions where ETH and MBL are clearly well identified, and a crossover region with a right shift (i.e., towards larger disorder) of the finite-size estimate of the transition point with system sizes.

TABLE I. Finite-size scaling results with single-size training, as a function of truncation order $N_c$.

| Truncation | $h_c$ | $\nu$ | $\chi^2$/dof |
|---|---|---|---|
| $N_c = 64$ | $3.16 \pm 0.13$ | $0.23 \pm 0.07$ | 0.03 |
| $N_c = 128$ | $3.19 \pm 0.09$ | $0.22 \pm 0.06$ | 0.13 |
| $N_c = 256$ | $3.25 \pm 0.09$ | $0.23 \pm 0.05$ | 0.32 |

*Finite-size scaling.* We define the finite-size pseudo critical point $h_c(L)$ as the disorder for which the fraction $f$ of MBL-classified samples equals 0.5. The finite-size scaling results for different truncation order $N_c = 64$, 128, 256 are summarized in Table I. In practice, we approximate the fraction $f$ by a cubic polynomial around the putative $h_c(L)$ fitted in the interval $[h_c(L) - w; h_c(L) + w]$ with $w = 0.6$ (giving the smallest error bars).

The scaling procedure leads to a critical disorder value $h_c \simeq 3.2$ that is lower than the usual estimate around $h_c \simeq 3.7$ [32], and extremely small values of $\nu \simeq 0.22$, which appear unreasonable. The underestimation of the critical disorder seemingly comes from the truncation preprocessing step, indeed $h_c$ increases as $N_c$ increases. Note that we needed to take aside $L = 22$ data for $N_c = 64$ (otherwise having $\chi^2$/dof $=$ 0.95): the number of truncated probabilities is too small to get a meaningful result for the largest size (see inset of Fig. 4).

*Understanding the black box: internal parameters of the network.* The most straightforward way to understand what the NN learnt is to directly look at their weights. Figure 5 shows a typical family of weights obtained after training on $L = 18$ data. The neurons split up in two symmetric groups: (i) [respectively, (ii)] half of the neurons weigh positively (respectively, negatively) the largest probabilities $p_i$ (corresponding to the smallest input indices) until input index $i \simeq 40$ then the next inputs are weighed negatively (respectively, positively). We observed that category (i) corresponds to neurons that activate most for an MBL-labeled sample, thus we denote them MBL detectors. Likewise, category (ii) is responsible for the detection of ETH features.

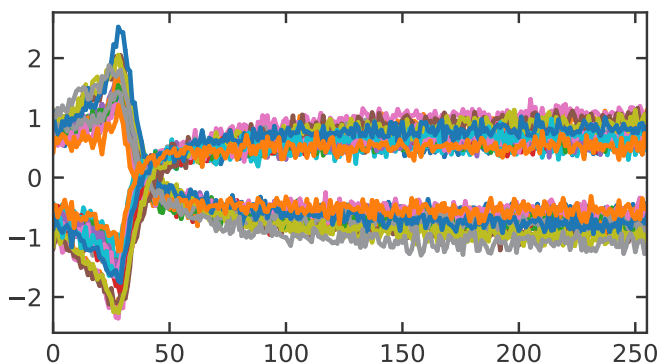

FIG. 5. Weights of the first hidden layer (32 neurons) of a typical training instance of a NN trained on $L = 18$ data for $N_c = 256$. Each color corresponds to one hidden neuron, its weights are connected to the input layer and are plotted against input neuron index.

Figure 5 points towards the relevance of the participation entropies $S_q^P$ for high values of $q$ (as the largest $p_i$ are more weighted by the NN), as a feature to classify the phases and detect the transition. The particular relevance of the IPR ($q = 2$) was also noted in the support vector machine analysis of a MBL transition in Ref. [26].

*Discussion.* One limitation of this setup is the possibility that a NN trained on a given $L$ could learn (i.e., reproduce the features of) a certain physical observable different from the one learned for a NN trained at a different $L$. Indeed, learning a certain classification model depends for instance on the NN capacity (number of layers/hidden neurons) relative to the complexity of the training dataset (that varies from one system size to another). Even more dramatically, Ref. [75] showed that different physical observables are learned depending on the amount of regularization, though this happened with support vector machines.

In addition, we find that a NN trained on a given system size in fact captures a model specific to this size. This can be seen for instance in a principal component analysis of the network weights (see Fig. 7 and its discussion in Sec. V). It has already been noticed that size-dependent features can indeed be captured [29,75]. It seems then illusory to achieve meaningful transfer learning like detecting the transition on $L_1$ data from a model trained on $L_2 \neq L_1$ data. In the next section, we present a solution aimed at addressing these two issues.

## V. MULTIPLE SYSTEM SIZE TRAINING

Most neural network architectures require having input data of fixed size. This comes from the fact that any fully connected layer needs a fixed number of ingoing connections. The chosen formatting of input data (Sec. III A) with fixed size allows us to use one unique NN to treat data from different system sizes on equal footing. Including all system sizes in the training dataset can be viewed as a regularization setup that prevents detection of size-specific features. Also, we hope that this will help the neural network to capture size-invariant features, i.e., features in the thermodynamic limit, in particular close to criticality.

In the following, we investigate what a neural network trained on a dataset containing system sizes $L = 16, 18, 20, 22$ all at once can learn and compare the results to the previous analysis (we refrain from using $L = 24$ data as not enough samples are available for training). To do so, we need to work at constant truncation order $N_c$ whatever system size is picked for training. The dataset has the same size as in the previous section, taking one fourth of samples from $L = 16$ data, one fourth from $L = 18$ and so on.

Figure 6 shows the fraction of MBL-classified samples defined in the previous section and displays similarities with Fig. 4 regarding the existence of fully-ETH and fully-MBL regimes located at the same regions. Nevertheless a striking

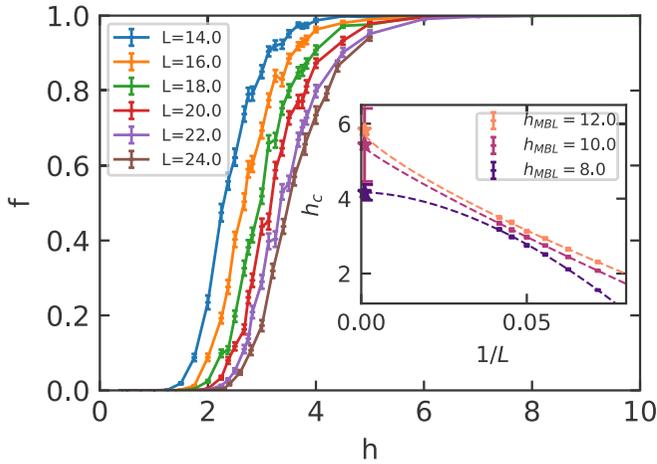

FIG. 6. Fraction of MBL-classified samples as a function of disorder strength for a NN trained on multiple system sizes all at once and evaluated on different system sizes. Predictions are averaged over 250 disorder realizations per disorder (with 100 eigenstates per realization) and 50 training instances. Truncation order is $N_c = 256$. The error bars indicate the statistical error due to sampling disorder realizations. Inset: finite-size scaling analysis with $h_c(L)$ defined as $f(h_c(L)) = 0.5$ for ETH-labeled data at $h_{\text{ETH}} = 0.25$ and MBL-labeled data at $h_{\text{MBL}} = 8.0$, $10.0$, $12.0$, the error bars on the final estimates come from the fitting procedure.

asymmetry from single-size training appears: a broadening of the curves in the crossover region.

Note that the figure above features nontrivial transfer learning: a neural network trained on $L = 16$, $18$, $20$, $22$ is asked to classify samples from system sizes $L = 14$ and $24$ for which it has never seen any samples before. This highlights one advantage of this multi-size training setup, namely its reduced computational cost. It is indeed reduced by a factor proportional to the number of considered system sizes and number of retrainings, which can represent a huge saving in computation time.

*Finite-size scaling, and dependence on training region.* We perform a finite-size scaling with varying training datasets which include MBL-labelled samples drawn from different disorder strengths $h_{\text{MBL}} = 8.0, 10.0, 12.0$ while the ETH-labelled samples are all taken from $h_{\text{ETH}} = 0.25$, because we noticed negligible change in the scaling for $h_{\text{ETH}} = 0.5$ or $1.0$. The results are summarized in Table II.

TABLE II. Finite-size scaling results with multiple-size training, for different values of the training disorder used to label the MBL phase. "Averaged" refers to the method defined in Sec. IV, "Individual" is defined in note [76]. Truncation order is $N_c = 256$.

| Data | Method | $h_c$ | $\nu$ | $\chi^2/\text{dof}$ |
|---|---|---|---|---|
| $h_{\text{MBL}} = 8.0$ | Averaged | $4.19 \pm 0.23$ | $0.57 \pm 0.08$ | 0.24 |
| $h_{\text{MBL}} = 8.0$ | Individual | $4.17 \pm 0.04$ | $0.58 \pm 0.01$ | – |
| $h_{\text{MBL}} = 10.0$ | Averaged | $5.43 \pm 0.98$ | $1.14 \pm 0.41$ | 0.04 |
| $h_{\text{MBL}} = 10.0$ | Individual | $4.93 \pm 0.10$ | $0.93 \pm 0.04$ | – |
| $h_{\text{MBL}} = 12.0$ | Averaged | $5.80 \pm 1.43$ | $1.29 \pm 0.64$ | 0.18 |
| $h_{\text{MBL}} = 12.0$ | Individual | $5.74 \pm 0.21$ | $1.27 \pm 0.09$ | – |

We found that including predictions obtained by transfer learning at $L = 14$ and $24$ system sizes considerably improve the results, in the sense that the fitting procedure converges with rather small error bars on $h_c$ and $\nu$. If $L = 14$ is taken aside, the error bars are multiplied by a factor of 4 and the fits do not converge if no transfer learning is done (performing the fit only on $L = 16, 18, 20, 22$).

If we apply the same fitting procedure as in previous section (tagged by "Averaged" in Table II), we obtain relatively large error bars. We were able to get the same estimates $h_c$ and $\nu$ but with error bars reduced by 10 using individual prediction of the critical point by each network (see the procedure detailed in note [76]).

The finite-size scaling analysis with varying training datasets leads to a somewhat unexpected result: whereas it is generally considered that the $h > 8$ region contains only strongly MBL eigenstates with very similar physical properties, still the neural networks learn different models resulting in estimates of $h_c$ ranging from $h_c \simeq 4$ to $h_c \simeq 6$, higher than the estimated value, and $\nu$ ranging from 0.5 to 1.5. This phenomenon can be rationalized with the following naive argument: samples in the transition region will be classified MBL for lower disorders if the MBL-labelled samples are themselves taken from region closer to the transition, thus shifting the transition point towards lower critical disorder. One can speculate that this finding actually echoes the nonuniversal multifractal properties of the MBL phase recently noticed in Ref. [33] and based on the same type of input data. Indeed, one can associate a different multifractal dimension (decreasing with $h$) to every $h_{\text{MBL}}$: the $h_{\text{MBL}}$ dependence could then be viewed as the manifestation of the varying multifractality in the MBL phase.

To circumvent this issue, one may for instance include samples from a range of disorder values all at once. However, we noticed that if we provide a training dataset containing MBL-samples from $h_{\text{MBL}} = 8, 10, 12$, the NN tend to capture $h_{\text{MBL}}$-averaged features of the dataset (see next paragraph), i.e., leading to predictions similar to those of a NN trained at $h_{\text{MBL}} = 10$.

*Analysis of network internal parameters.* The two previous training setups—single and multiple system size training—give different critical estimates. We now try to understand the source of these differences using a principal component analysis (PCA). The use of PCA has already proven to be useful in many previous works (see, e.g., Refs. [20,77]). It is used here as a dimensional reduction procedure and allows to represent the weights connected to one hidden neuron (a $N_c = 256$-component vector) as a point in Fig. 7 after a projection onto the two principal components. Even though PCA suffers from limitations [71] and more advanced techniques exist [78], we found in our case that 90% of the total variance is accounted by these two principal components. This already allows us to reach informative conclusions from this simple PCA approach.

Figure 7 confirms many points discussed previously. The weights split up into two symmetric categories by a sign change, this corresponds to the MBL and ETH detectors revealed in Fig. 5, visible here through the PCA representation. It also confirms that single-size training on $L$ data leads to capturing $L$-specific features. A hierarchy appears where the weights corresponding to training at a given system size $L$ are next to the weights for $L \pm 2$.

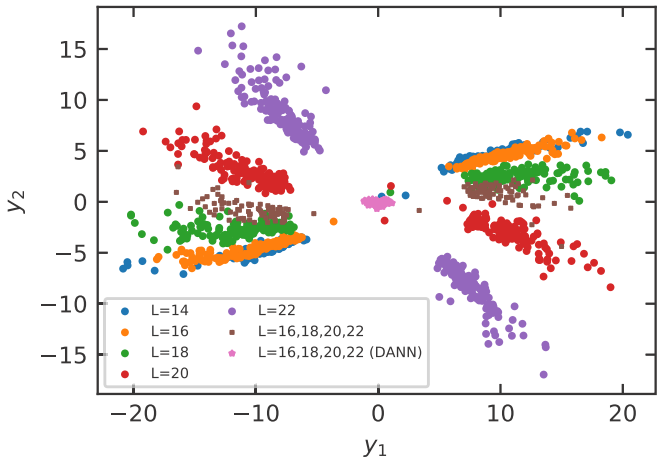

FIG. 7. PCA representation of the weights learned after training on single system size datasets (Sec. IV), multiple system size datasets (Sec. V), supplemented by an $L$-adversarial component (Sec. VI). Each dot is a weight projected on the two principal axis of the PCA analysis (which accounts for 90% of the total variance). Five training instances are included for each training case.

However, the weights learned in the multiple system size training case overlap the weights of the NN trained at $L = 18$ and 20, seemingly capturing an averaged model among the system sizes $L = 16, 18, 20, 22$. This shows that the NN does not actually capture size-independent features (which would manifest by a uniform distribution of weights over the $L$-specific subspace of weights) but rather in a weaker way, it uncovers averaged features that are shared by all the provided system sizes. To corroborate this point, we trained a NN on system sizes $L = 14, 16, 18, 20$ and we also notice the same averaging behavior, i.e., this time the NN captured features similar to those captured for $L = 16$ and 18 trainings.

*Discussion.* Previous section showed that if one neural network is trained only on one system size $L$, the respective predictions of the $L$-specific NN cannot necessarily be comparable, hence rendering any finite-size-scaling procedure questionable. The multiple-size training setup was expected to produce more reliable predictions, but the results are somewhat disappointing for different reasons.

First, the obtained error bars on $h_c$ and $\nu$ are higher than in the previous case (for the same fitting procedure), pointing to the fact that the definition of $h_c(L)$ may not be the most suitable choice in this setup. One can for example define $h_c(L)$ as the disorder for which all samples are MBL, i.e., when the fraction $f$ first reaches 1: this would affect, not dramatically but in a sensible way, the final estimates of $h_c$ and $\nu$. This difference in treating the ETH and MBL phases could be justified by the various physical observations that the MBL transition displays asymmetries: see for instance the avalanche scenario which implies that a thermal bubble can more easily destabilize a MBL sample than a MBL bubble does for an ETH sample [79], as well as that the critical point is localized [80]. However, this is in our opinion a too strong bias and would go against our original goal of providing as minimal physical input as possible. Second, the transition point greatly depends on the region of the phase diagram used for training (this was also noticed in Refs. [6,40]). This is clearly a limitation of our setup since one would want the critical parameters to be insensitive to the location of the training data in the phase diagram. Third, the analysis of the weights revealed that this setup leads to the learning of an averaged model of the system sizes provided in the dataset. Next section aims at circumventing these limitations, in particular by introducing a constrained setup that is designed to prevent the NN from capturing size-dependent features or size-averaged behaviors.

## VI. SYSTEM SIZE ADVERSARIAL TRAINING

The two previous sections pointed out the difficulty to fight against dataset dependence of the NN predictions. The best that we could obtain with the preceding architectures is a NN that has captured averaged features of the training dataset when it contains data from multiple system sizes. We recall that our objective was to use a diverse dataset to expose the NN to rather different samples labeled the same to later achieve good generalization either to the transition region or even to unseen system sizes (for $L = 24$, for instance, since it becomes increasingly hard to generate a large amount of training data).

Domain-adversarial neural networks (DANN) have been introduced in Ref. [81] in order to tackle domain adaptation, i.e., when the datasets at training and test/prediction time come from similar but different distributions. The general principle is to learn features that cannot discriminate between the training (source) and test (target) domains. In practice, this is achieved through an adversarial setup that promotes the emergence of features that are (i) discriminative for the main learning task on the source domain and (ii) indiscriminate with respect to the shift between the domains. This idea has been recently used in two works [27,77] dealing with phase classification, where the source domain consisted of the extremal region of the phase diagram and the target domain being the transition region.

Expanding on these ideas, we exploit the specificity of this scheme to force the NN to learn features that are insensitive to the system sizes it has been trained on. In other words, the goal is to use DANN to learn feature detectors that are $L$-invariant. In particular, a DANN contains two supplementary components shown in Fig. 8: a system size classifier and a gradient reversal layer. The latter component is the only nonstandard part in this architecture and works by leaving the input unchanged during forward-propagation and reverses the gradient by multiplying it by a negative scalar during the back-propagation: this results in changing the sign of the gradient of the feature extractor parameters with respect to size classifier loss. That way, the common feature extractor is adjusted to make the task of the phase classifier as easy as possible while making that of the system size classifier as hard as possible. If the network reaches equilibrium, the selected features are the best suitable to identify which phase a sample lies in, while containing no information about which system size it emanates from.

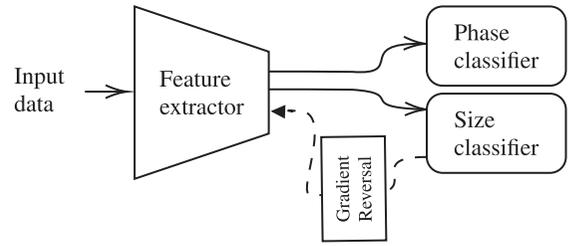

FIG. 8. Neural network containing an adversarial component applied on the system size label.

*Learning L-invariant features*

In this section, we study the predictions of a DANN trained on data from system sizes $L = 16, 18, 20, 22$ all at once. In particular, we analyze the effect of the adversarial component compared to the setup of the previous section. The feature extractor part (see Fig. 8) is kept identical from previous sections (i.e., same hyperparameters, NN structure, etc.). The system size classifier consists of 4 softmax neurons corresponding to each provided system size and outputs which can be interpreted as the probability of a sample to be from any of the given system size. The loss function contains now an additional term that takes care of the size labels (second term in the following equation):

$$\mathcal{L} = \sum_{x, \mathbf{y}^h, \mathbf{y}^L} \underbrace{\sum_{j=1}^{2} y_j^h \ln\left(f_j^h(x)\right)}_{\text{Phase classifier loss}} + \underbrace{\sum_{j=1}^{4} y_j^L \ln\left(f_j^L(x)\right)}_{\text{Size classifier loss}}, \quad (7)$$

where $\mathbf{y}^h$ (respectively, $\mathbf{y}^L$) is the two-dimensional (respectively, four-dimensional) one-hot vector representing the phase label (respectively, the system size label) of sample $x$, $\mathbf{f}^h$ (respectively, $\mathbf{f}^L$) is the corresponding two-dimensional (respectively, four-dimensional) softmax output of the phase classifier (respectively, the system size classifier). Because of the adversarial component, the optimization process will keep the size classifier loss at much higher values (in practice orders of magnitude larger) than the phase classifier loss: the NN will thus be discriminative for the phase classification task and indiscriminate with respect to the shift between the $L$-data domains.

Adversarial learning is generally considered to be a hard task [82], for instance nonconvergence can occur with oscillations of the optimized parameters. Training is known to be very sensitive to the hyperparameter selections since any unbalance between the two adversaries can lead to overfitting or other unwanted phenomena. In particular, we noticed that the weights of the feature extractor tended to take arbitrarily large values (increasing with training time). This has the effect of increasing the variance of the predictions from one training instance to another and may also cause overfitting.

Therefore we found it crucial to add a $L^2$ weight decay term in the cost function (7), in the form $\mu |W|^2$ with $W$ being the internal parameters of the feature extractor. This regularization technique requires however a good choice of $\mu$. If $\mu$ is too large, the constraint is too strong and the optimization procedure struggles to minimize the classifier losses. If $\mu$ is too small, the limitations presented above are not corrected, i.e., the model variance stays high. After fine-tuning, we found that $\mu = 0.05$ gives good results. We checked that the finite-size scaling of previous section with the same regularization (weight decay with $\mu = 0.05$) gives same critical values with no better error bars.

TABLE III. Finite-size scaling results using a DANN approach for multiple-size training, as a function of the training disorder used to label the MBL phase.

| Training data | Method | $h_c$ | $\nu$ | $\chi^2$/dof |
|---|---|---|---|---|
| $h_{\text{MBL}} = 8.0$ | Averaged | $5.50 \pm 1.02$ | $1.18 \pm 0.40$ | 0.21 |
| $h_{\text{MBL}} = 8.0$ | Individual | $5.38 \pm 0.14$ | $1.14 \pm 0.06$ | – |
| $h_{\text{MBL}} = 10.0$ | Averaged | $5.44 \pm 1.05$ | $1.18 \pm 0.44$ | 0.13 |
| $h_{\text{MBL}} = 10.0$ | Individual | $5.75 \pm 0.19$ | $1.37 \pm 0.08$ | – |
| $h_{\text{MBL}} = 12.0$ | Averaged | $5.63 \pm 1.21$ | $1.25 \pm 0.52$ | 0.01 |
| $h_{\text{MBL}} = 12.0$ | Individual | $6.16 \pm 0.34$ | $1.54 \pm 0.15$ | – |

*a. Finite-size scaling.* We perform the finite-size analysis of the NN predictions as before. The predictions for $L = 14$ and 24 are obtained by transfer learning. The results are summarized in Table III.

We notice various improvements from last section. The adversarial component helps reducing the training region dependence noticed before (also noted in Ref. [27]). The critical disorder $h_c \simeq 5.5$–6 is still higher than the conventional estimate and $\nu \simeq 1.2$ is also (slightly) higher. In addition, the error bars are increased, especially for $h_{\text{MBL}} = 8.0$ and 10.0. The second fitting method (see Ref. [76]) is also moderately less effective due to a higher variance between each neural network variance (see Appendix B for a detailed discussion).

We stress that these results cannot be compared to the ones obtained in Ref. [27] because the setup used there differs in several ways: whole eigenfunctions (up to size $L = 18$) and single-size training are used, and the adversarial component is used differently to reduce the discrepancy between samples from the strongly ETH/MBL regions and from the intermediate region.

*Interpretation of the network parameters.* The PCA representation of the DANN weights in Fig. 7 shows that this setup allows some apparent independence of the model with respect to system size, indeed, the weights are homogeneously distributed over $L$-specific weights subspaces.

Similarly to Fig. 5 showing the weights connecting the input layer to the first hidden layer, Fig. 9 shows the weights connecting the feature extractor to the size classifier. We argue that the $L$ invariance of the features is achieved by reaching the following trivial equilibrium configuration: any feature vector (output of the feature detector) is multiplied by the weight vector $\mathbf{W}_L$ towards the $L$ classifier with $L = 16, 18, 20, 22$ and Fig. 9 shows precisely that $\mathbf{W}_{L=16} = \mathbf{W}_{L=18} = \ldots$ Due to softmax normalization of the system size classifier, this leads to a $L$ classification of any sample of equal probability of belonging to any of the provided system sizes.

*Discussion.* This setup has proved to improve many limitations of the previously considered architectures, namely reducing the training dataset as well as the training region dependences. Nevertheless, we found that training a DANN is very sensitive to hyperparameters choices (regularization parameter $\mu$, etc.) and chosen NN structure (depth, etc.), hence requiring very good calibration otherwise instabilities can rapidly occur. We also noticed greater variance of the predictions from one instance to another (see Appendix B).

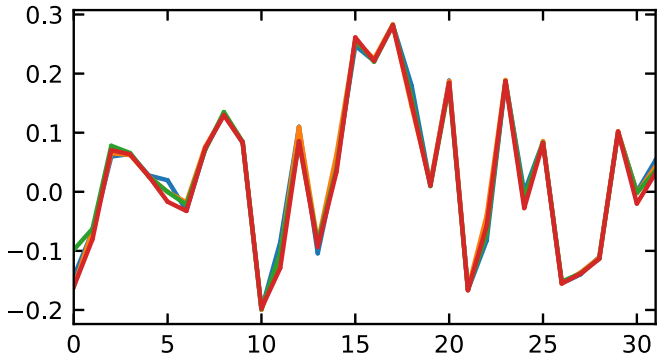

FIG. 9. Weights connecting the feature extractor to the system-size classifier plotted against hidden layer neuron index of a training instance of a DANN trained on $L = 16, 18, 20, 22$ data for $N_c = 256$. Each color corresponds to one of the 4 size-classifier neurons.

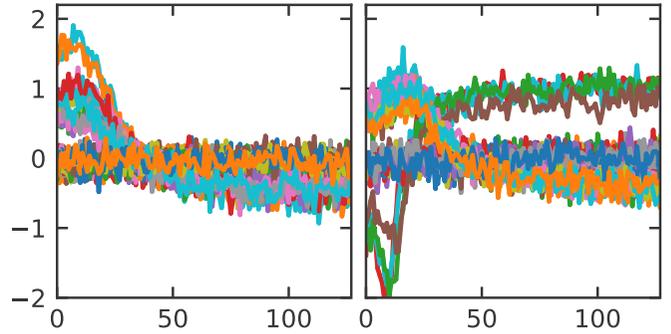

FIG. 10. Weights of the first hidden layer (32 neurons) of two exemplary training instances of a NN trained on $L = 18$ data with ReLU activation units. Each color corresponds to one hidden neuron, its weights are connected to the input layer (here having $N_c = 128$ components) and are plotted against input neuron index.

## VII. DISCUSSION OF RESULTS

The initial goal of this work was to attempt a finite-size study of model (1) using neural networks. Our analysis revealed numerous difficulties: the scaling procedure appeared very sensitive to the neural network hyperparameters (the specific choice of activation function, the addition of dropout or weight decay), as well as the imposed structure (whether an adversarial component is added or not). In addition to that, there is no inherent criterion that allows us to discriminate these different external choices, and as a matter of fact, we can consider our analysis as a kind of *model exploration* (different machines with the same accuracy have different ways of solving the same task) rather than *model selection* (selecting the machine that achieves the highest accuracy on a given task).

The limitations also arose from the dependence on the particular choice of training dataset, we highlighted that the NN predictions and ultimately the finite-size scaling actually depend on the region of the phase diagram used for training. Moreover when the training dataset includes data from several system sizes, the NN tend to extract average features that do not permit accurate transfer learning. Including a constraint to fight against this behavior (here in the form of $L$-invariant adversarial component) improves the situation to a certain extent at the cost of having to fine-tune extra hyperparameters and thus potentially adding more bias in the final estimates.

These limitations occurred even though we provided the best possible input data (i) giving directly the wave functions with a controlled compression step and (ii) also in terms of available system size (up to $L = 24$ in the MBL context). Nevertheless we find that multi-size training of NN allows to grasp consistent finite-size trends based on a limited amount of disorder realizations. This points towards one of the NN advantages, that is its reduced computational cost compared to conventional methods. Another interesting point (discussed in Appendix B) which we discovered in investigating the contributions to the variance of the prediction is that the network output correlates quite well with the entanglement entropy.

The finite-size scaling led to critical values of $h_c$ and $\nu$ always larger than conventional estimates: $h_c$ being around $\simeq 5$–6 while $\nu$ is about $\simeq 1.2$–1.5. The finite-size scaling of the MBL transition in model (1) (with random disorder) has been shown to be particularly difficult, with system sizes available from exact diagonalization argued to be too small to probe the correct criticality [62]. We do not find that the machine learning analysis improves this situation, at least within the setup and input data that we chose. In particular there is no obvious reason to trust more the neural networks final results (again within the approach chosen in this work) than the ones reached within the conventional approach. The generic trend that seems to emerge is towards a larger extent of the ETH phase, even though we emphasize that no critical field $h_c(L)$ (obtained for a single system size $L$) exceeds the value $h_c \simeq 3.7$ reached from the conventional approach within error bars. This last remark could mean that the finite-size scaling (and thus final estimate) could be improved if one would be able to improve on uncertainties (coming from multiple sources) on individual $h_c(L)$ obtained from the NN analysis.

Our thorough finite-size study of this phase transition leads to the conclusion that one always has to be aware of the multiple bias that can possibly arise when using neural networks and its power might be limited to qualitative predictions rather than precise estimations, here for instance finite-size scaling. This is particularly relevant for phase transitions whose nature or universality class is unknown or debated and/or for which the input data has some limitations (e.g., in terms of the range of size accessible in our case).

We finish with suggestions on possible improvements of this situation. In the case of the MBL transition studied here, one can certainly improve the quality of the output by providing more physical knowledge of the transition in the input data (such as when using the entanglement spectrum). Alternatively, one could keep the same generic input data (wave-function coefficients) but use recent results [33] on the finite-size scaling of participation entropies to try to build an improved network architecture as well as to better interpret the outputs. For the more generic case of unknown phase transition, further work is needed to ascertain the reliability of finite-size scaling within the neural network approach, ideally providing tools to construct and understand a generic finite-size scaling theory of the network prediction. Recent works [83–87] connecting the renormalization group and the neural network construction may be first steps in this direction.

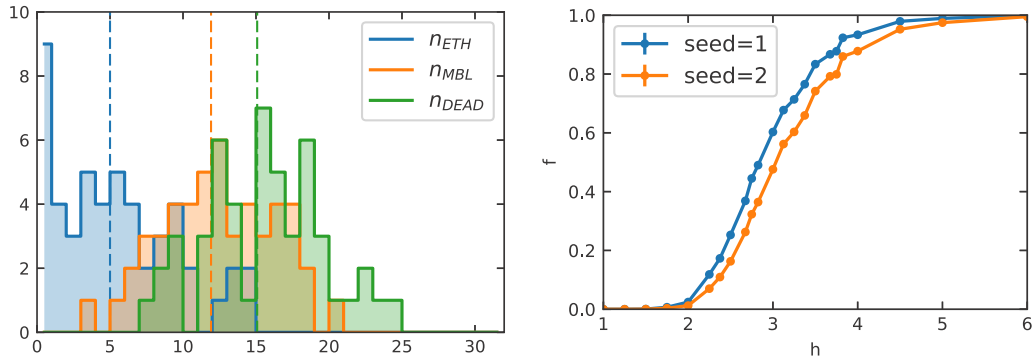

FIG. 11. (Left) Histograms of the number of ETH detectors (blue), MBL detectors (orange) and dead neurons (green) per training instance (having 32 hidden neurons) calculated over 50 NN instances. (Right) Fraction of MBL-classified samples of the two NN instances shown in Fig. 10, averaged over 250 disorder realizations per disorder (and 100 eigenstates per realization).

### ACKNOWLEDGMENTS

We thank Patrick Huembeli and Alexandre Dauphin for introducing us to the domain adaptation thematics as well as Evert van Nieuwenburg, Nicolas Macé, and Nicolas Laflorencie for very useful comments. This work is supported by a grant from the Fondation CFM pour la Recherche, and benefited from the support of the project THERMOLOC ANR-16-CE30-0023-02 of the French National Research Agency (ANR) and by the French Programme Investissements d'Avenir under the program ANR-11-IDEX-0002-02, reference ANR-10-LABX-0037-NEXT. We acknowledge PRACE for awarding access to HLRS's Hazel Hen computer based in Stuttgart, Germany under Grant No. 2016153659, as well as the use of HPC resources from CALMIP (Grants No. 2018-P0677 and No. 2019-P0677) and GENCI (Grant No. 2018-A0030500225). Our shift-invert [70] numerical calculations are based on the linear algebra libraries: PETSC [88,89], SLEPC [90], and STRUMPACK [91,92]. The neural network calculations are performed with TENSORFLOW [93].

## APPENDIX A: HOW RELU ACTIVATION FUNCTIONS INDUCE BIASES IN THE ANALYSIS

ReLU activation functions are broadly used in the machine learning community as well as in many of its applications to physics [3,27,35]. The main motivation comes from the fact that they do not suffer from saturation contrary to their sigmoid or tanh counterparts. However it is known that training with ReLU units can lead to dead neurons, i.e., neurons that output zero whatever input value comes in. Although this phenomenon effectively allows to learn sparser representations, in our case it drastically reduces the NN capacity to a point such that the underfitting regime is actually reached.

Figure 10 reveals the existence of a third category of neurons: dead neurons that have zero weights for all incoming connections, which are invisible in Fig. 5 of the main text where we used ELU activation functions. The appearance of such neurons comes along with great variability from one NN instance to another, some instances having more MBL or ETH detectors than others. This is visible in the histograms of the weights in the left panel of Fig. 11 that shows that there is on average 37% of MBL detectors, 16% of ETH detectors, and 47% of dead neurons from statistics of 50 training instances. The right panel of Fig. 11 indeed shows how a variable ratio of MBL/ETH detectors shift the transition point and therefore add a bias that is only due to the NN structure.

Furthermore, we find that the addition of dropout into the NN with Relu activation is very problematic due to the phenomenon shown in Fig. 12: if one follows the NN output of individual samples during training, dropout induces huge variations. This can be explained by the unbalance between the number of ETH and MBL detectors, indeed dropping ETH detectors will greatly impact the classification since they already are less numerous on average than MBL detectors. In addition, one necessarily has to stop training at some step and these great variations prevent any choice of stopping criterion. With ELU units on the other hand, we observed that there is on average the same number of MBL and ETH detectors, thus dropping randomly detectors does not impact on average the classification.

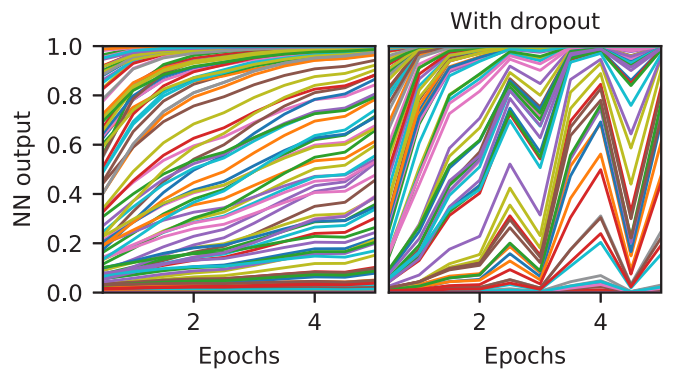

FIG. 12. NN output of 200 eigenstates (different colors) plotted against training steps (left) without dropout and (right) with dropout at $h = 3.0$, for a NN trained on $L = 16$ data.

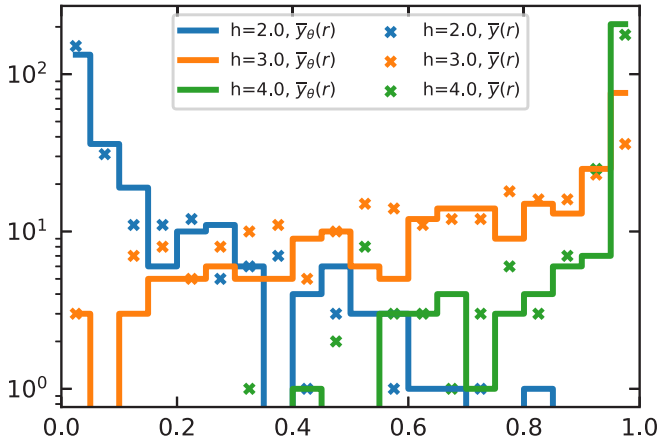

FIG. 13. Histogram of $\bar{y}_\theta(r)$ and $\bar{y}(r)$ as defined in the main text obtained with 250 disorder realizations. Predictions are obtained from $L = 18$ data and 50 NN trained according to the DANN setup.

## APPENDIX B: SOURCES OF CLASSIFICATION VARIANCE, PREDICTION FOR INVIDUAL EIGENSTATE

As noted in Refs. [35,36], the NN approach allows for a direct low-resolution analysis of the transition, i.e., at the level of eigenstates. In this Appendix, we highlight several interesting features based on the analysis of prediction for individual eigenstate.

*a. Eigenstate-to-eigenstate, sample-to-sample variance.* First, we consider variations of classification from one disorder realization to another. For a given neural network $\theta$, we study the distribution of classifications across disorder realizations. As done in the main text, we average the classification (0 meaning ETH, 1 MBL) of individual eigenstates sharing the same disorder realization $r$ denoted as $\bar{y}_\theta(r)$. Figure 13 shows an histogram $\bar{y}_\theta(r)$ for a typical NN $\theta$ for 250 disorder realizations. We have checked that this picture is stable for all training instances and for any of the considered setups.

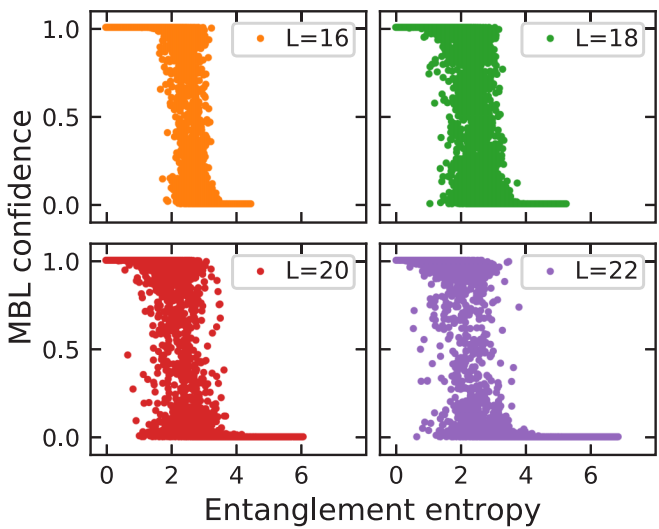

FIG. 14. MBL raw confidence as a function of entanglement entropy for 100 individual eigenstates of 100 different disorder realizations at $h = 3.0$, for different system sizes in the multi-size training setup.

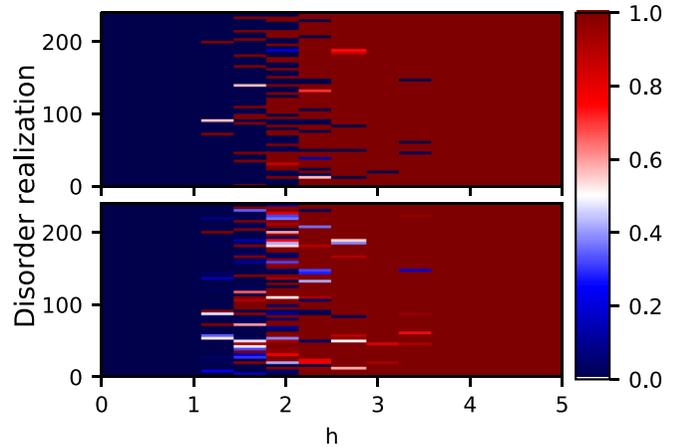

FIG. 15. Average of one eigenstate per disorder realization over 50 training instances ($\bar{y}_r$ as defined in the main text) as a function of disorder strength and realization number for (top) a multi-size training setup and (bottom) a DANN setup with predictions obtained at $L = 18$.

For disorder strengths slightly lower (respectively, higher), i.e., at $h = 2.0$ (respectively, $h = 4.0$) than the crossover point (here around $h = 3.0$ for $L = 18$), the distribution is peaked around 0 (respectively, 1) meaning that almost all eigenstates from any disorder realizations are classified as ETH (respectively, MBL). More interestingly, for $h = 3.0$ the NN detects both ETH and MBL eigenstates within the same disorder realization. This effect is discussed in the next paragraph.

Figure 13 also shows the distribution over 250 disorder realizations of the average classification of eigenstates sharing the same disorder realization and 50 independent training instances denoted by $\bar{y}(r)$. We chose results for the setup with adversarial component which displays the most variance from one training instance to the other (this is quantified in next paragraph). The distribution roughly follows the distribution of $\bar{y}_\theta(r)$ meaning that the same physical picture explained above persists for all training instances on average.

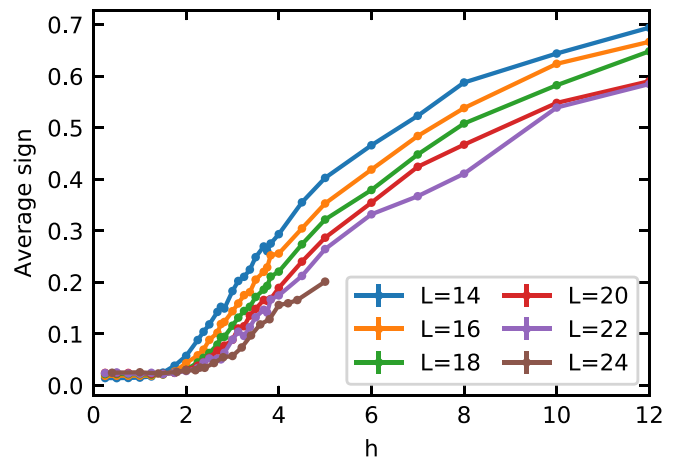

FIG. 16. Average sign averaged over 200 disorder realizations per disorder for different system sizes.

TABLE IV. Finite-size scaling results when inputing signed coefficients $c_i$ to the NN.

| Setup | Training data | Method | $h_c$ | $\nu$ | $\chi^2$/dof |
|---|---|---|---|---|---|
| Multisize | $h_{\text{MBL}} = 12.0$ | Averaged | $6.87 \pm 2.28$ | $1.77 \pm 0.92$ | 0.11 |
| Multisize | $h_{\text{MBL}} = 12.0$ | Individual | $6.88 \pm 0.33$ | $1.77 \pm 0.13$ | – |
| DANN | $h_{\text{MBL}} = 12.0$ | Averaged | $7.86 \pm 3.83$ | $2.24 \pm 1.64$ | 0.07 |
| DANN | $h_{\text{MBL}} = 12.0$ | Individual | $7.87 \pm 0.57$ | $2.25 \pm 0.24$ | – |

*b. Correlation of individual eigenstate prediction with its entanglement entropy.* The fact that, close to the transition, the network predicts both ETH and MBL eigenstates in the same disorder realization at the same energy density is reminiscent of what was observed in Ref. [94], where a bimodal distribution of entanglement entropy was observed also at the individual disorder realization level close to the transition. This suggests looking at the correlation between the prediction of each eigenstate and its entanglement entropy.

This correlation is represented in Fig. 14 for four different sizes in the multi-size training setup for $h = 3.0$. We clearly see that eigenstates with low (high) entanglement are systematically classified as MBL (ETH) and maximizing (minimizing) the MBL confidence to be 1 (0). In agreement with Ref. [94], we have checked that an important number of disorder realizations contain at the same time eigenstates with low and high entanglement (and correspondingly high and low MBL confidence). For each system size, there exists a crossover region for intermediate values of entanglement entropy for which the full range of MBL confidence can be found. This gives rise to the increased variance of the prediction near the transition region, and most certainly to the higher error bars observed there.

*c. NN variance.* As pointed out in the main text, we observed the largest model variance in the DANN setup (Sec. VI). To show this, we pick one eigenstate per disorder realization and compute its average classification over 50 training instances denoted by $\bar{y}_r$. We do the same for the other 250 different disorder realizations and the result is showed in Fig. 15 as a function $h$ for both multisize training and $L$-adversarial training.

For the multisize training, there is almost no variance: all NN classify the same eigenstate almost identically, which can be observed as predictions are most of the time close to 1 or 0 in the left panel of Fig. 15. For the DANN setup, the fluctuations due to disorder realizations are supplemented by fluctuations due to NN classifications. In effect, Fig. 15 shows that a given eigenstate can sometimes be classified as ETH and MBL for two different training instances, with averages more often closer to intermediate values ∼0.5. Figure 15 also allows to detect disorder realizations for which the average prediction is markedly different from others for a given strength of disorder. Quite interestingly, the NN predictions presents a certain asymmetry in the transition, with more MBL-classified samples on the ETH side than ETH samples on the MBL side.

### APPENDIX C: WORKING WITH AMPLITUDE $c_i$

For a given eigenstate $|n\rangle = \sum_i c_i |i\rangle$, in the main text, we chose to provide the probabilities $p_i \equiv |c_i|^2$ as input to the NN. We investigate here whether restoring the signs, i.e., taking directly the amplitude $c_i$ as input data, would allow for a better estimate of the transition. Indeed, this input contains more information than contained in $p_i$ input, which could potentially lead to less biased finite-size estimates. Note that as the Hamiltonian in Eq. (1) is real and symmetric, all $c_i$ are real (up to degeneracies which can occur only exceptionally due to the random part in the Hamiltonian). As before, we keep only the $N_c$ highest amplitudes $c_i$ (sorted by their absolute value). For illustration, we show in Fig. 16 the average sign defined as

$$\overline{\text{sign}(|n\rangle)} = \frac{\left| \sum_{i=1}^{N_c} \text{sign}(c_i) |c_i|^2 \right|}{\sum_{i=1}^{N_c} |c_i|^2} \quad \text{(C1)}$$

for different system sizes. For small disorder, the average sign stays small, close to zero. We indeed expect eigenfunctions coefficients to be Gaussian distributed around zero in the ETH phase. As disorder strength increases, the average sign grows more rapidly for lower system sizes until it eventually reaches 1 in the high-disorder limit. Again, we understand this limit very well, as each eigenstate is dominated by a single coefficient in the $S^z$ basis in the infinite disorder limit.

We take the multisize training architecture and the same hyperparameters as in Secs. V and VI and attempt a finite-size scaling analysis. Results are summarized in Table IV. Despite inputing in principle more physical information, we find no noticeable improvement on the error bars of critical estimates, within the setup used.